\documentclass[a4paper]{article}
\usepackage{bm}
\usepackage{subfigure}
\usepackage{multicol}

\usepackage{mathtext}         
\usepackage[T2A]{fontenc}    
\usepackage[utf8]{inputenc}  
\usepackage[russian,english]{babel}   

\usepackage{amssymb}                                          
\usepackage{amsmath}                                          
\usepackage{amssymb}

\usepackage{graphics}
\usepackage{graphicx}
\usepackage{wrapfig,epsfig,epsf}
\usepackage{bm}    

\makeatletter
\renewcommand\@biblabel[1]{#1.} 
\makeatother

\usepackage[left=2.5cm,right=2.5cm,top=4cm,bottom=2.7cm]{geometry}

\def\beq#1{\begin{equation}\label{#1}}
\def\eeq{\end{equation}}
\def\beqa#1{\begin{eqnarray}\label{#1}}
\def\eeqa{\end{eqnarray}}

\def\myfrac#1#2{\left(\frac{#1}{#2}\right)}

\def\comment#1{\relax}

\begin{document}

\title{Fast Radio Bursts}

\author{S.B. Popov$^1$, K.A. Postnov$^{1,2,3}$, M.S. Pshirkov$^{1,4,5}$\\
$^1$Sternberg Astronomical Institute,\\
Lomonosov Moscow State University,\\
119234, Russia, Moscow, Universitetskii pr., 13\\
$^2$ Kazan Federal University, Kazan, Kremlyovskaya st., 18\\
$^3$ National Research University Higher School of Economics, \\ Myasnitskaya st., 20, 101000 Moscow, Russia\\
$^4$ Institute of Nuclear Research RAS, \\ 60-letiya Oktyabrya pr., 7a, 117312 Moscow, Russia\\
$^5$ Astro-Space Center, Lebedev Physical Institute,\\
Pushchino Radio Astronomical Observatory, \\ 142290, Pushchino, Moscow Region, Russia\\ 
}

\maketitle

\section*{Abstract}
The phenomenon of fast radio bursts (FRBs) was discovered in 2007. These are powerful (0.1-100 Jy\footnote{Jansky (Jy) is the unit of spectral flux density used in radio astronomy.  1 Jy = $10^{-26}$~W~m$^{-2}$~Hz$^{-1}$.)}) single radio pulses with durations of several milliseconds, large dispersion measures, and record high brightness temperatures suggesting coherent emission mechanism. As of time of writing, 32 FRBs were recorded. There is also one repeating source from which 
already hundreds of bursts have been detected. The FRB rate is estimated to amount up to several thousand per day over the sky, and their isotropic sky distribution likely suggests a cosmological origin. Since the discovery, different hypotheses on the possible FRB nature have been proposed, however up to now the origin of these transient events remains obscure. The most prospective models treat them as being related to a bursting emission from magnetars -- neutron stars whose activity is due to their magnetic field dissipation, -- or as being analogs of giant  pulses observed from several radio pulsars . Future increase in the statistics of the observed FRBs and an improvement upon the FRB population characteristics will make possible to use them as a new tool to probe the intergalactic medium and to test fundamental physical theories.  

\section{Introduction}

Transient (i.e. rapidly emerging for a relatively short time) cosmic electromagnetic sources provide a wealth of information about astrophysical objects. 
Their   observations at different frequencies constitute an important part of modern astrophysical researches. The transients can be related to well-known objects (for example, flaring stars), can be repeating (for example, giant pulses of pulsars), or can be associated with unique events (for example, supernova explosions or binary neutron star coalescences). The detection efficiency of transient phenomena, clearly, depends on the sensitivity of a given detector, its field of view, the duration of the transient, as well as on the background level and possible interferences.  

In different electromagnetic bands, astronomical observations of transients have their own peculiarities. In the radio band, which will be mostly discussed in this review\footnote{This is an updated version of the review accepted for publication in Physics Uspekhi (2018).}, a lot of different types of transients have been observed (see, e.g., the review  \cite{2004NewAR..48.1459C}).
Some of them remains unidentified with other astronomical sources and their nature remains obscure. For example, the source  GCRT J1745-3009 in the Galactic center \cite{2005Natur.434...50H, 2010ApJ...712L...5R} which  demonstrates outbursts, typically lasting minutes, with an observed flux of $\sim$Jy. There are longer transients (for example, flares of active galactic nuclei), and there are much faster ones.

\begin{figure}
\center{\includegraphics[width=150mm]{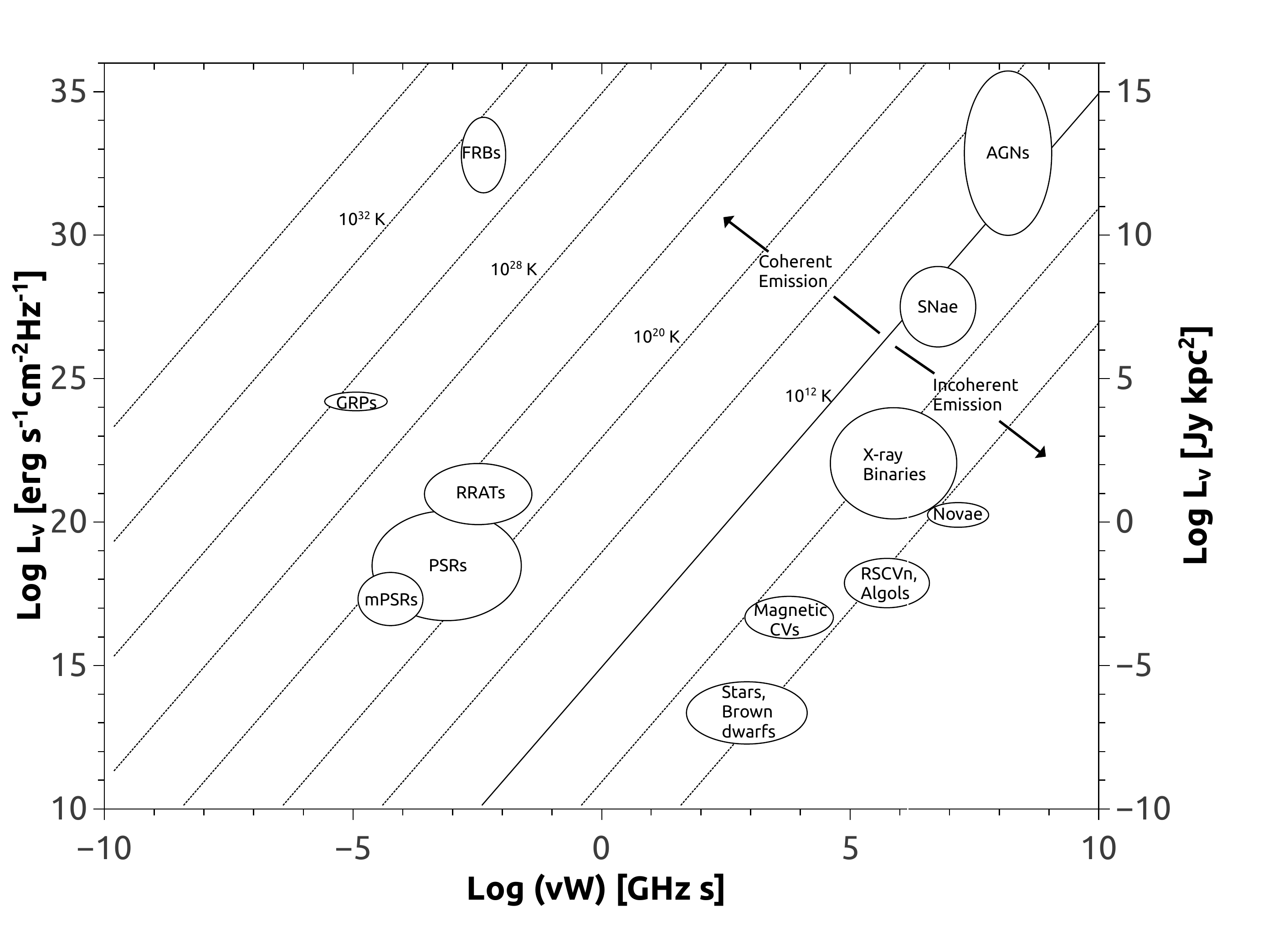}}
\caption{Radio transients on the $L_\nu$~--~$\nu W$ plane. 
The straight lines show constant brightness temperature
$T_b$. Shown are pulsars (PSR), rotating radio transients (RRATs), millisecond pulsars (mPSR), giant pulses from pulsars (GRP), 
active galactic nuclei (AGN), supernovae (SN), X-ray binaries (X-ray binaries), RS CVn and Algol stars (RSCVs, Algols), magnetic cataclysmic variables (magnetic CVs), normal (Main sequence) stars and brown dwarfs.
}
\label{lnuw}
\end{figure}

Here we will consider fast transient radio flares observed by modern radio telescopes at frequencies from hundred MHz to several GHz.
From a large variety of radio transients (see, for example, 
the review \cite{2004NewAR..48.1459C}), the most interesting at present are the so-called 'fast radio bursts' (FRBs) discovered in archive data of the Parkes Radio Sky Survey in 2007  \cite{2007Sci...318..777L}. 
It is natural that their unusual properties, such as   short durations $W\sim 1-10$~ms,  high radiation intensities (a peak spectral flux density $S_\nu$ up to hundred Jy),  large dispersion measures \footnote{The dispersion measure (DM) is the integral of the electron number density along the line of sight to the source
$DM=\int n_e dl$. DM is measured in units of 
[cm$^{-3}$~pc], were the distance 1 parsec (pc) $\approx$ 206265 a.u. $\approx 3\times 10^{16}$~m.} $DM\sim 500$~cm$^{-3}$~pc, and  a high sky event rate of the order of several thousand per day  \cite{2016MNRAS.460L..30C} arouse much interest. 

\subsection{Coherent emission of FRBs}

One of the distinct features of FRBs is their large dispersion measure significantly exceeding the dispersion measure due to the Galactic plasma in the given direction. This suggests large distances to these sources, most likely indicating  their extragalactic nature. This conclusion is also supported by isotropic sky distribution of FRBs (as long as this can be inferred from the small source statistics and inhomogeneous sky coverage in different radio surveys). The high intensity and short duration of FRBs suggest a high brightness temperature of radio emission pointing to a coherent radiation mechanism. In the Rayleigh-Jeans limit  
($h\nu\ll kT$), the brightness temperature $T_b$ is 
\beq{}
kT_b=I_\nu \frac{c^2}{2\nu^2},
\eeq  
where $I_\nu$ [W~m$^{-2}$~Hz$^{-1}$~sr$^{-1}$] is the radiation intensity, $c$ is the speed of light, $\nu$ is the frequency,
$k$ is the Boltzmann constant. 
By assuming the angular size of the source
$\theta$ and the burst duration casually bounded, 
$W\sim  l/c$, where $l=D\theta$ is the transversal size of the source at distance $D$, we obtain the brightness temperature estimate:
\beq{}
2\pi kT_b=\frac{S_\nu D^2}{(W\nu)^2}
\eeq 
or, by inserting the characteristic values, 
\beq{}
T_b\approx 10^{35.8}[\mbox{K}] \myfrac{S_\nu}{1~\mbox{Jy}}\myfrac{(D/1~\mbox{Gpc})}{(\nu/1~\mbox{GHz})(W/1~\mbox{ms})}^2.
\eeq
(Here all quantities are given in the observer's rest frame). 
If the source is moving relativistically, this expression for the brightness temperature in the proper source's rest frame should be reduced by factor $\sim \gamma$, where  
$\gamma\gg 1 $ is the Lorentz-factor of the emitting region.
\footnote{From relativistic kinematics, when transforming from  the laboratory frame (K) to the rest frame (K') we have 
$I_\nu = I'_{\nu'} 
{\cal D}^3$, $\nu=\nu' {\cal D}$, ${\cal D}=\frac{1}{\gamma(1-\beta\mu)}$ is the Doppler factor, $\gamma=\frac{1}{\sqrt{1-\beta^2}}$ is the Lorentz-factor,  $\mu=\cos\theta$ is cosine of the angle between the velocity vector and the direction to the observer in K.
For $\gamma \gg 1$ we have ${\cal D}\approx \gamma$, and from (1) we find $T_b\approx T_b'\gamma$. Another, more phenomenological derivation of this result can be found in \cite{2014PhRvD..89j3009K}. } 
These estimates show that even for FRBs at Galactic distances (kiloparsecs), the brightness temperature will be definitely higher than
$10^{12}$~K.

Following \cite{2004NewAR..48.1459C, 2015aska.confE..51F},
we present different populations of radio transients in the $\nu W$--$L_\nu$ plane, where
$W$ is the pulse width, and $L_\nu$ is the luminosity per unit frequency
(see Fig. \ref{lnuw}). 
In radio astronomy, one frequently uses the 'pseudoluminosity' $L_\mathrm{psd,\nu}=S_\nu D^2$ obtained from the observed flux and the source distance estimate, which is commonly identified with the specific luminosity $L_\nu$. We follow this tradition to display sources in the Fig.  \ref{lnuw}. 

At brightness temperatures above $\sim 10^{12}$~K the radiation from cosmic sources should be coherent because at higher values the electrons would rapidly cool down due to Compton losses \cite{1969ApJ...155L..71K}. The radio
emission from pulsars, rotating radio transients (RRATs), and FRBs belongs to this class. The radio emission from some X-ray binaries, nova stars, flaring stars and brown dwarfs, as well as from most of supernovae is thermal and non-coherent in most cases. 

\subsection{FRB power}

The energetics of FRBs is also among their important characteristics. By estimating the specific radio luminosity of a source at distance 
$D$ from the observed spectral flux density 
$S_\nu$, assuming radiation into the solid angle $\Delta \Omega$,  $L_\nu=4\pi D^2 S_\nu (\Delta\Omega/4\pi)$, and setting the energy released in the burst equal to 
$\Delta E =\nu L_\nu W$, we find 
\beq{e:DeltaE}
\Delta E \sim \nu S_\nu W 4\pi D^2\myfrac{\Delta \Omega}{4\pi}=10^{39}[\hbox{erg}] \myfrac{\nu}{1~\mbox{GHz}}\myfrac{S_\nu}{1 ~\mbox{Jy}}\myfrac{W}{1~\mbox{ms}}\myfrac{D}{1~\mbox{Gpc}}^2.
\eeq
The relativistic motion of the source decreases this value in the source rest frame $K'$ by the factor ${\cal{D}}^3\simeq \gamma^3$ 
(assuming isotropic emission in $K'$ so that $(\Delta\Omega'/4\pi)=1$). 
For sources at cosmological distances, the energy release in a short coherent burst requires radiation mechanisms in relativistic plasma
(see below in Section \ref{s:models}). For sources at Galactic distances ($\sim$10 kpc) the energy output $\sim 10^{29}$~ergs is comparable to 
the characteristic energetics  of stellar flares, which was used as a ground for a Galactic model of FRBs \cite{2014MNRAS.439L..46L}.

\subsection{Specific of short burst detections}

The propagation of a radio pulse in the ionized interstellar medium (ISM) or in the intergalactic medium leads to the broadening of the detected pulse
(see \cite{2003ApJ...596.1142C} for more detail). At first, the pulse widens due to the frequency dependence of the refraction index of the ionized medium  (the dispersion measure effect). For a cold plasma, 
$n=\sqrt{1-\myfrac{\omega}{\omega_p}^2}$, where $\omega_p$ is the plasma 
frequency. The pulse time delay at different frequencies is determined by the dispersion measure DM:
\beq{}
t_{\mathrm{DM}}\approx 4.2 \times 10^3~ [\mu\mbox{s}] \mathrm{DM}\myfrac{\nu}{1\mbox{GHz}}^{-2}\,.
\eeq
For a non-coherent (channel) dispersion in the receiver band 
$\Delta \nu$ at frequency $\nu$ the pulse width would be   
\beq{}
\Delta t_{DM} \approx 8.3 ~[\mu \mbox{s}] \mathrm{DM} \frac{(\Delta \nu/1~ \mbox{MHz})}{(\nu/1~ \mbox{GHz})^3}\,.
\eeq  

Second, for an optimal registration, the signal is 'dedispersed', i.e. the time lag at different frequencies due to DM is removed, therefore the 
inevitable error in DM, $\delta \mathrm{DM}$,
additionally increases the pulse width by about 
$\Delta_{\delta\mathrm{DM}}=\Delta t_{DM}(\delta\mathrm{DM}/\mathrm{DM})$. 
Third, the pulse width increases because of a finite receiver bandwidth, $\Delta t_{\Delta\nu}\sim \Delta\nu^{-1}$. Finally, 
the pulse width increases by an amount 
$\tau_d$ due to the scattering on density fluctuations of the turbulent interstellar and intergalactic medium (the scintillation effect). By the present time, the last effect was observed only in some FRBs; for other sources, only upper limits are known. 
Like the dispersion measure time lag, the pulse width can increase along the entire path from the source to the observer. A significant enhancement of the FRB pulse broadening over the Galactic pulsars supports their extragalactic origin. The existing data do not allow us to uniquely determine where the pulse widening occurs -- either near the source in its host galaxy, or somewhere between the source and the observer, though the  former scenario seems to be more plausible
\cite{2016ApJ...818...19K,2016ApJ...832..199X,2017ApJ...835...29Y,2018arXiv180305697F}. 
The measured pulse scattering width 
$\tau_d$ strongly varies from source to source and can amount up to several msec 
at the frequency 1 GHz. For several FRBs, the frequency dependence of this broadening was also estimated, 
$\tau_d\sim\nu^{\alpha}$, where $\alpha\sim-4$. 
The fast growth of the pulse width caused by the scattering can be also a reason preventing FRB observations at low frequencies (see below).

The total pulse width increase is approximately equal to 
\beq{}
\Delta t=(\Delta t_{\mathrm{DM}}^2+\Delta t_{\delta\mathrm{DM}}^2+\Delta t_{\Delta\nu}^2+\tau_d^2)^{1/2}\,.
\eeq

Unlike the scattering effects, the interstellar dispersion effects can be partially or fully removed by a dedicated signal processing. 
As in searching for transients the dispersion measure is not known in advance, it is chosen such that the signal dispersion after the dedispersion procedure is minimal, 
$\Delta t\to \Delta t_0=(2\Delta t_{\mathrm{DM,min}}+\tau_d^2)^{1/2}$. 

After the dedispersion, pulses with amplitude exceeding some threshold signal-to-noise ratio (SNR) are searched for in the records. As the duration of the pulse, $W_i$, is unknown, its value is chosen to maximize SNR. For the root-mean-square (rms) noise amplitude 
$\sigma_n$ and the pulse amplitude $S_i$ 
(or its fluence $F_i\approx S_i W_i$),
the optimal SNR is 
\beq{}
\mbox{SNR}_i=\myfrac{(F_i/W_i)}{\sigma_n}\sqrt{\frac{W_i}{W_n}},
\eeq
where $W_n$ is the radiometer noise correlation time. As in the radio dishes $\sigma_n\simeq S_{n}/\sqrt{\Delta \nu W_n}$, 
where $S_{n}$ is the antenna noise in Janskys, 
the optimal signal-to-noise ratio does not depend on 
$W_n$, $\mbox{SNR}_i\propto F_i/\sqrt{W_i}$, and the narrower the pulse at a given fluence, the higher its detection SNR. Of course, for a large fluence (strong signal) even wider pulses can be easily detected. 

A radio pulse propagating in the ISM gets wider, as discussed above, with its fluence being kept constant. If the pulse width due to the propagation and detection effects exceeds the intrinsic pulse width, $W_b > W_i$, the optimal SNR changes correspondingly as 
$\mbox{SNR}_b=\mbox{SNR}_i\sqrt{W_i/W_b}$. Clearly, the pulse widening due to propagation in ISM significantly complicates the signal registration.

Here we would like to make one physical note. For radio pulses with a high brightness temperature, the induced scattering effects in the surrounding medium (Compton scattering on electrons and Raman scattering on plasmons) can be significant. These effects were investigated in paper 
\cite{2008ApJ...682.1443L}. It was shown that during propagation of a single pulse in a medium, the optical depth is mainly determined by its width $\Delta t$, and the induced scattering is important only in plasma near the source (for example, in the pre-supernova stellar wind in the case of radio emission from gamma-ray bursts). The induced scattering effects inside the source itself can be neglected if the pulse is generated in a relativistically moving plasma. This made possible to obtain the lower limit on the Lorentz-factor of plasma inside FRB, 
$\gamma>(3-4)\times 10^3$, which is relatively insensitive to the model parameters. Note that this condition is also met both for giant radio pulses generated in pulsar magnetospheres and for the magnetar model of FRBs discussed below in Section \ref{s:magnetar_model}, and significantly 
restricts (in addition to other arguments) the possibility of Galactic generation of FRBs in stellar flares  \cite{2016ApJ...818...74L}. 

Despite the huge interest to the FRB phenomenon and increasing number of papers devoted to the FRB problem, their nature remains unknown. Frequently (and correctly) this situation is compared to the initial studies of cosmic gamma-ray bursts (GRBs) in the 1970s -- mid 1990s, until the first optical identification of GRB in a remote galaxy firmly established its extragalactic origin (see, e.g.,  the review \cite{1999PhyU...42..469P}).
Therefore, until the distance to FRBs is measured from astronomical observations, a plethora of possible physical models remains possible. In what follow, after a brief discussion of the history of the FRB studies, we will consider the main phenomenological properties of FRBs as a new astronomical phenomenon  
(Sec. \ref{s:Observations}), with a separate description of the presently unique repeating FRB 121102 (Sec. \ref{s:FRB121102}). Then in Sec. 4 we 
will discuss different scenarios proposed for explanation of FRBs, in more detail we present two the most likely extragalactic FRB models: the FRB model as a non-coherent collection of nanosecond giant pulses from young pulsars  (Sec. \ref{s:GP_model})
and the model of synchrotron maser emission during giant bursts from magnetars (Sec. \ref{s:magnetar_model}). 
In Section 4.4 we discuss extragalactic FRBs as possible intergalactic medium probes and their cosmological implications. In the Conclusion section, the prospects of the FRB studies will be briefly summarized.

\section{History of FRBs discovery}

In this Section, we present a brief review of the most important episodes in the short history of early FRB studies.

\subsection{Key episodes of the early stages of FRB studies}

In some sense, the discovery of Rotating  Radio Transients (RRATs) 
\cite{2006Natur.439..817M} can be considered as a precursor to the FRB discovery. It was necessary to develop a complicated technique, as described in \cite{2003ApJ...596.1142C}, to identify single short radio flares. In the case of RRATs, the analysis found a periodicity in records from a given source. Thus,  this allowed the identification of RRATs as a sub-group of radio pulsars.

Presently, more than hundred RRATs are known. The mechanism of generation of these short intense radio bursts
has not been identified as yet. However, studies of the activity of these sources allowed to show their affinity to radio pulsars as sources with extreme nulling (i.e. temporarily absence of any radio emission for many successive spin periods) exceeding 95\% of  time \cite{2013IAUS..291...95B}.

It is important to stress that RRATs have been discovered (and more and more sources are continuously found) via single pulse searches, i.e. not due to searches of periodic radio emission, as in the case of radio pulsars. Therefore, just in the beginning of the 21st century, an effective method of robust identification of single millisecond radio bursts was developed. This opened the way for the discovery of an absolutely new phenomenon -- fast radio bursts.

The first FRB discovery was announced in the autumn 2007 \cite{2007Sci...318..777L}. The burst itself was observed in 2001, and later on was named FRB 010724 (year-month-day). It was discovered by the 64-meter radio telescope in Parkes (Australia) during a radio pulsar survey at the frequency 1.4 GHz. The observed flare position was three degrees off the Small Magellanic Cloud. The signal demonstrated a very large peak flux exceeding 30 Jy and a short duration of $<5$~ms. 
The key feature of this burst was its huge dispersion measure: $DM=375$~cm$^{-3}$~pc. This was significantly larger than the Milky Way contribution in this direction (see the standard model of the electron density distribution in \cite{2002astro.ph..7156C} and a new model in \cite{2012MNRAS.427..664S}). Follow-up observations lasted 90 hours, but no new flares were detected. As the event was discovered during a survey, it was possible to make an estimate of the rate of such events in the sky. Assuming that the dispersion measure was due to the contribution from the intergalactic medium, the authors obtained  the rate estimate: 90 bursts per day in the fiducial volume 1~Gpc$^3$. This roughly corresponds to thousands of bursts every day in the whole sky.
It is interesting to note that after the discovery of more than thirty FRBs this rough estimate is still valid within a factor of a few.

Soon after the first observational publication on FRBs, several theoretical studies appeared. In some of them ideas already proposed in the discovery paper \cite{2007Sci...318..777L} were developed, in others --- new hypotheses were put forward.
Still, without new observational data, the topic was not actively studied. 

The detection  of the second burst, FRB010621, was announced only in 2012 \cite{2012MNRAS.425L..71K}. The burst was observed in one of the outer beams of the Parkes telescope. The width of the pulse was 7.8 ms and no consequent flares were found.  Also, this burst was nearly two orders of magnitude weaker than the first one and was found close to the Galactic plane. Therefore, the event looked much different from the first Lorimer burst. Taking into account the problems of perytons (see the following subsection), it was not obvious whether these two events represented a new class of astrophysical phenomena. Archival searches of records from different radio telescopes (in the first place those working at lower frequencies) gave null results.

The breakthrough happened thanks to the paper \cite{2013Sci...341...53T} where the authors reported discoveries of four new millisecond radio transients with large dispersion measures at high Galactic latitudes. This publication can be considered as a starting point of the modern history of FRBs.

\subsection{Perytons}

In the short history of FRBs there was a notable episode related to the discovery of the so-called perytons.
This type of short radio bursts was identified \cite{2011ApJ...727...18B} soon after the publication of the paper by Lorimer et al. \cite{2007Sci...318..777L}. As in the case of FRBs, the discovery of these events was done due to the analysis of archive records of the Parkes telescope.

In the first paper about the new type of radio transients \cite{2011ApJ...727...18B} (where also was proposed the popular name of  this phenomenon), the authors presented data about 16 events. All of them had durations about ten milliseconds. In contrast to FRBs, perytons were recorded in all (or many) beams of the Parkes telescope diagram. These events did not seem to be uniformly distributed in time on the day or/and year time scales. They mostly happened during working hours late in the morning. All perytons had similar spectral characteristics. 
Signals were delayed at lower frequencies similar to the dispersion effect in the interstellar (or intergalactic) medium. The frequency dependence of the time shift $\delta t $ was very similar to the classical value $\delta t \sim \nu^{-2}$.
However, for some events significant variations were observed. 
Some bursts could come in series (for example, once 11 flares were detected in less than five minutes).

Already the authors of \cite{2011ApJ...727...18B} proposed that perytons were not of space origin, and most probably are technogenic. Note that at that time only one FRB (the Lorimer burst) was known. Formally, for many perytons it was possible to determine the dispersion measure. The obtained values were about 300-400 cm$^{-3}$~pc (see Fig. 9 in  \cite{2014ApJ...797...70K}), which is extremely close to the DM of the Lorimer burst -- 375 cm$^{-3}$~pc. Thus, the origin of the only known FRB could have been questioned: was it an astrophysical phenomenon, or it had a terrestrial (even technogenic) origin, like perytons?

Perytons have been searched by several radio telescopes (Arecibo, C. Jansky VLA, Allen Telescope Array), also new observations were made at Parkes (see a review  and references in \cite{2014ApJ...797...70K}). However, only at the Bleien observatory in Switzerland similar signals were detected \cite{2012MNRAS.420..271K}.

The possible origin of perytons was analyzed in details in several papers (see \cite{2014ApJ...788...34K},
\cite{2014ApJ...797...70K} and references therein). 
Many different hypotheses were discussed: atmospheric discharges, meteors, and various technogenic processes (e.g., a transit of an airplane in the field of view of a radio telescope). Still, it was not possible to pinpoint a single model that could explain all observed features of perytons.

The solution came out unexpectedly. In December 2014, a new equipment for monitoring RFI (radio frequency interference) has been installed at Parkes.  Then, a very detailed study to uncover the origin of perytons was performed. This included new observations and data mining \cite{2015MNRAS.451.3933P}. Simultaneously with Parkes, observations were performed with the Australia Telescope Compact Array (ATCA) and in India using the Giant Meterwave Radio Telescope (GMRT). Only at Parkes perytons were detected, and it became clear that this was a local problem. Also, in addition to a signal at 1.4 GHz, the monitoring system detected the emission at 2.3-2.5 GHz, which seemed to have a technogenic origin.

A careful analysis (and experiments!) indicated that perytons appeared from microwave ovens at the observatory! For a particular orientation of the telescope, when the door of an oven was opened while it was still working, a peryton was detected at the frequency $\sim$1.4 GHz.

Additional studies demonstrated that FRBs, known at that time (including the Lorimer burst),  do not have peculiarities typical for perytons. Thus, the astrophysical origin of FRBs seemed to be robust. However, the exact nature of these sources remained unknown.

\section{Recent results of observations and phenomenology of FRBs}
\label{s:Observations}

Modern FRB observations started with the paper by Thornton et al. \cite{2013Sci...341...53T} reporting four new events found in the archival Parkes records. As of time of writing (spring 2018), more than 30 FRBs are known plus one repeating source. 
The online catalog of FRBs is available at the web-site frbcat.org  \cite{2016PASA...33...45P}. Brief reviews dedicated to FRBs have been published (see   \cite{2017JApA...38...55R}, \cite{2017arXiv170902189P}, \cite{2018arXiv180409092K}, and references therein).

\subsection{Non-repeating bursts}

\begin{figure}
\center{\includegraphics[width=150 mm]{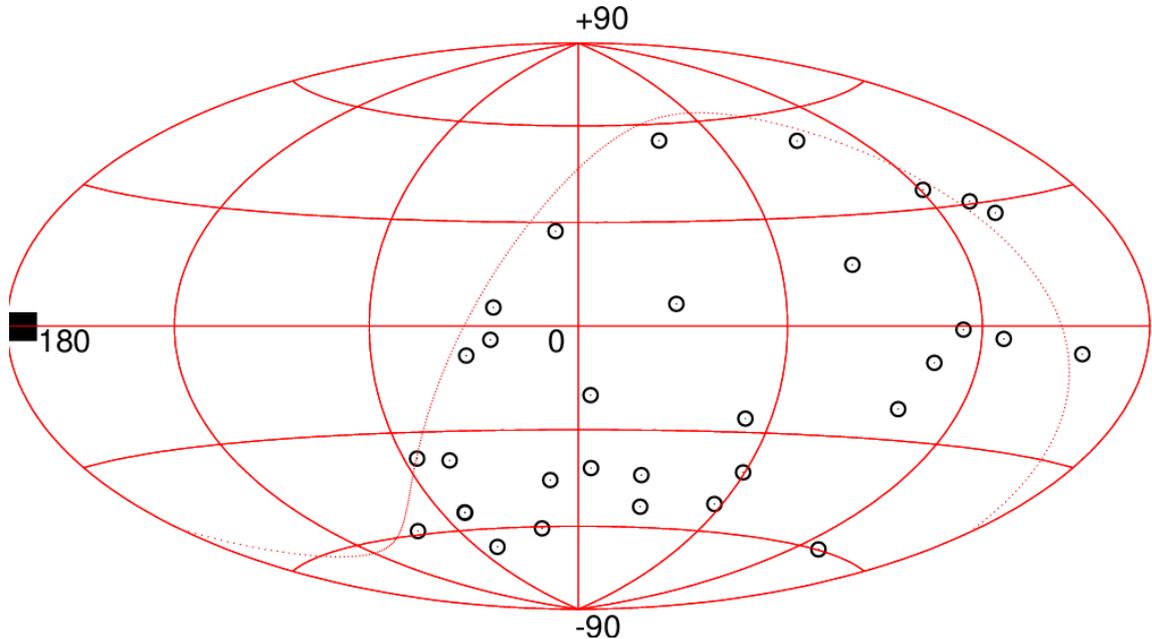}}
\caption{ The sky distribution of known FRBs in the Galactic coordinates (latitude increases from right to left). The position of the only known repeating source --  FRB121102, -- is marked with a solid square. The dotted (red) line is the celestial equator. As most of FRBs are detected by the Australian instruments, the north-south asymmetry is visible. Data taken from the online catalog frbcat.org.
}
\label{map}
\end{figure}

Presently, more than 30 single (non-repeating) bursts are known.
Most of them (25 events) were detected with the 64-meter telescope at Parkes. Five FRBs are detected by UTMOST (Australia). Finally, ASKAP (Australia) \cite{2017ApJ...841L..12B} and the Green Bank Telescope (GBT) (USA) detected one FRB each. New sources are announced approximately once per month. The sky distribution of known sources is shown in Fig. \ref{map}.

The typical peak flux of FRBs is $\sim 1$~Jy. However, for bright events (e.g., FRB010724, FRB170827) it reaches tens of Jy, and for extreme cases, like FRB150807, it is $>100$~Jy!

Pulse width of detected events lies in the range from a fraction of millisecond up to $\sim30$~ms. Also, in some cases a microstructure is visible inside a narrow pulse. For example, a very bright event recently detected by UTMOST -- FRB170827, -- had a pulse width of $\sim 0.4$ ms and demonstrated three subcomponents, the most narrow with a duration of only 30 $\mu$s \cite{2018arXiv180305697F}. Subpulses of similar duration were also observed in the repeating source FRB121102 \cite{2018Natur.553..182M}.

The sky distribution of known FRBs is significantly biased because nearly all of them were detected by the Australian instruments (Parkes, UTMOST, ASKAP). Also it is necessary to note that many FRBs were discovered in archival  pulsar survey data, i.e. the sky coverage was not uniform. However, the statistical analysis does not reject the hypothesis that the sky distribution of known FRBs is uniform.

Dispersion measures of FRBs lie in the range from $\sim170$ up to  $\sim 2600$ cm$^{-3}$~pc. At first, a dedicated search for events with large DM (few thousand cm$^{-3}$~pc) did not produce any results \cite{2016MNRAS.460.3370C}. However, recently, several bursts with $DM>1500$~cm$^{-3}$~pc were discovered. The record belongs to FRB 160102 with $DM=2600$~cm$^{-3}$~pc \cite{2018MNRAS.475.1427B}.

As FRBs were mostly detected using archival data, it was impossible to look for transient counterparts at other wavelengths. The first burst discovered in the real time was FRB140514 \cite{2015MNRAS.447..246P}. 
This offered an opportunity to initiate a follow-up program
all over the electromagnetic spectrum. No related transients were found. This result allowed to reject (at least for the given source) all models related to supernova explosions, gamma-ray bursts, and some other bright transients which have been proposed as possible sources of FRBs. 

After FRB140514 several other real time detections were made: 
FRB150215 \cite{2017MNRAS.469.4465P}, FRB 150418 \cite{2016Natur.530..453K}, FRB 150807 \cite{2016Sci...354.1249R}, and then four events FRB 150610, FRB151206, FRB151230, FRB160102 presented in \cite{2018MNRAS.475.1427B}. In all cases there were no firm identifications of any counterparts. This also allowed many catastrophic scenarios of FRBs to be excluded.

In the case of FRB150418, it was proposed that the burst was related to a slow radio transient \cite{2016Natur.530..453K}. Based on this identification, a host galaxy was proposed for this FRB. However, later on it was demonstrated that two radio sources are unrelated \cite{2016ApJ...821L..22W, 2017MNRAS.465.2143J,2016A&A...593L..16G}.

Gamma-ray FRB counterparts have been actively searched for by {\it Fermi} space observatory \cite{2016MNRAS.460.2875Y, 2017ApJ...842L...8X}. In addition, a search for FRBs connected with GRBs (radio observations started after a GRB alert) was performed \cite{2014ApJ...790...63P}. No robust associations have been found. In one case (FRB131104), a candidate was proposed -- a gamma-ray burst with  a duration of several hundred of seconds \cite{2016ApJ...832L...1D}.  However, later the association between the FRB and the GRB was found to be unreliable \cite{2017JApA...38...55R}.

Presently, observations of non-repeating FRBs do not provide much information about spectral properties of these events. The bursts have been detected in relatively narrow bands either at the central frequency 1.4 GHz (Parkes, Arecibo), or $\sim 840 $ MHz (UTMOST, GBT). 
Intensive searches at lower frequencies with different telescopes (including LOFAR) gave null results 
\cite{2014A&A...570A..60C},

\cite{2016MNRAS.458.3506R},

\cite{2016MNRAS.458..718C},

\cite{2016ApJ...826..223B}.
In particular, in \cite{2016ApJ...826..223B}, based on non-detection of FRBs at low frequencies and assuming a power-law spectrum in a wide range of wavelength (from few centimeters to $\sim 1$ meter), the authors infer limits on the spectral index $\alpha$ ($F\sim \nu^\alpha$, where $F$ is the burst fluence): 
$-7.6<\alpha<5.8$. Nevertheless, considering all uncertainties, this result is not very constraining for FRB models.

Why FRBs are not detected at frequencies below 800 MHz is still unclear. It can be an intrinsic property of the emission mechanism. However, this is more likely to be due to absorption at low radio frequencies. In the first place, a major role can be played by the medium in the immediate surroundings of a  source (see, e.g. \cite{2016MNRAS.462..941L} and a more detailed and recent discussion in \cite{2018arXiv180401104P}). An analysis of different possibilities and predictions for future low-frequency observations (for example, with  such instruments as CHIME and HIRAX which will operate at frequencies  $\sim 600$  MHz) can be found in \cite{2017MNRAS.465.2286R}.

A recent search for FRBs at frequencies 300-400 MHz was conducted as part of the  Green Bank Northern Celestial Cap (GBNCC) pulsar survey 
\cite{2017ApJ...844..140C}. 
No FRBs have been found. The authors provide constraints on the spectral index under different assumptions about source properties and parameters of the medium along the line of sight.

No new events have been found during two-years (518 hours of observations in the period from July 2015 to August 2017) search in the project ALFABURST \cite{2018MNRAS.474.3847F}. However, this is not unexpected taking into account the small field of view of the telescope. The whole survey is equivalent to the observation of a 10-square degree field for one hour. For bright bursts an expected event rate in such a field is one per day. Hopefully, higher sensitivity of the Arecibo telescope could allow to detect  weaker bursts which could  have much higher event rate. Yet this null result does not enable important constraints on the FRB spectra, spatial distribution, or their luminosity function.

For several sources, the radio emission polarization was measured. The first case was FRB110523
\cite{2015Natur.528..523M}. The signal was linearly polarized at the 44\% level. The detection of a rotation measure of $RM=-186.1\pm1.4$~rad~m$^{-2}$ allowed estimation of the average magnetic field (weighted with the electron number density) along the line of sight: 0.38 $\mu$G. The analysis demonstrated that most probably the main contribution comes from the local medium in the vicinity of the source (contributions due to the intergalactic medium and interstellar medium in the Milky Way are relatively small).

Oppositely, FRB140514 demonstrated only circular polarization 
\cite{2015MNRAS.447..246P} at the 20\% level. Later on, three
other  linearly polarized sources were detected.  
In the case of  FRB150418, the degree of polarization was equal to 8.5\% \cite{2016Natur.530..453K}.  Such a low value did not enable any solid estimates of the magnetic field along the line of sight to be made. For FRB150215, the 40\% linear polarization was measured, but RM was found to be compatible with zero (within the uncertainty range) \cite{2017MNRAS.469.4465P}. Finally, a very high degree of polarization -- 80\%, -- was reported for FRB150807 \cite{2016Sci...354.1249R}. The relatively small value of the rotation measure, $RM\approx12$~rad m$^{-2}$, indicates that the medium around the source is not strongly magnetized. Correspondingly, this allowed to put constraints on parameters of the intergalactic magnetic field and turbulence. 
The repeating source FRB121102 is an outlier in this respect as well. It shows a nearly 100\% linear polarization and very large variable rotation measure: $RM\sim10^{5}$~rad~m$^{-2}$ \cite{2018Natur.553..182M}. More details on this source will be given below in Section 3.2.

Recently, in two cases, --- FRB151230 and FRB160102, --- the detection of both (circular and linear) polarizations was reported \cite{2018MNRAS.tmp.1076C}. For FRB151230 the polarization degree is not high, and the rotation measure is zero within the uncertainties. The circular polarization is even formally consistent with zero: 6$\pm$11\%. The linear polarization is 35$\pm$13\%. 
The radio emission from FRB160102 is significantly polarized. The linear polarization degree is even consistent with 100\%: 84$\pm$15\%. The circular polarization degree  is 30$\pm$11\%. Note that this source has the highest dispersion measure among the known FRBs, DM$\sim 2600$~cm$^{-3}$~pc, and a significant rotation measure RM$\sim 220$~rad~m$^{-2}$. Needless to say that RMs in addition to DMs can  be very useful probes of the medium in the FRB surroundings \cite{2018arXiv180401104P}.

Statistical properties of FRBs have been examined in many papers (see e.g., \cite{2017RAA....17....6L},
\cite{2016MNRAS.461..984O},
\cite{2017arXiv171008026R}, \cite{2018MNRAS.474.1900M},  \cite{2018ApJ...858....4N} and references therein). Not surprisingly, up to now the conclusions are very uncertain due to a small number of known events.

Estimates of the FRB rate on the sky are being continuously updated. They fall within the range from a few thousand up to a few ten thousand events over the sky sky per day for fluences above a few $\times0.1$~Jy~ms. The weakest sources detected at Parkes have fluences $\sim0.55$~~Jy~ms. Data are considered to be complete only for fluences above $\sim 2$~Jy~ms.

The authors of \cite{2017RAA....17....6L} give the following estimate:
\begin{equation}
dN/dF = (4.14\pm 1.3)\times 10^3\, \left(F/\mathrm{Jy}\,  \mathrm{ms}\right)^{-1.14\pm 0.2}\, \text{d}^{-1}. 
\end{equation}
In \cite{2017AJ....154..117L}, an estimate of 587 events per day for fluxes $>1$~Jy was given. A more detailed estimate (for different ranges of the Galactic latitude) is provided by the same authors in \cite{2016arXiv161200896V}.
In \cite{2018MNRAS.475.1427B}, the value $1.7\times10^3$ bursts per day for the whole sky was obtained based on the sample of FRBs detected at Parkes with fluences $>2$~Jy~ms.

\begin{figure}
\center{\includegraphics[width=120mm]{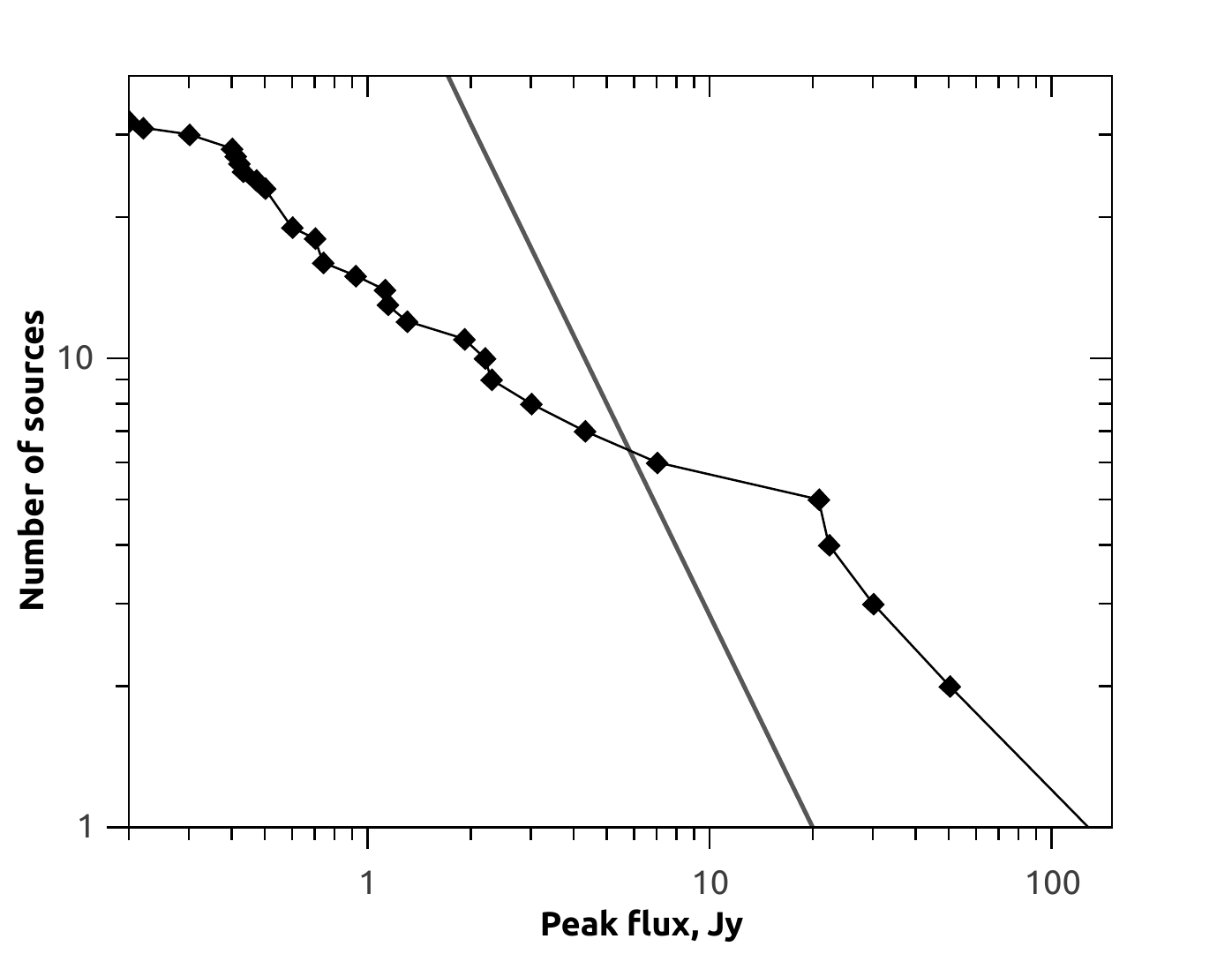}}
\caption{The integral peak flux distribution of fast radio bursts: Log $N(>S)$ -- Log $S_\mathrm{peak}$. The solid line corresponds to the law $N(>S)\sim S^{-3/2}$. Data from the on-line catalog frbcat.org.
}
\label{n_s32}
\end{figure}

In the zero approximation, we could assume that the peak flux distribution of FRBs (Log $N$ -- Log $S$) can be described by the law for the flat space: $N(>S)\sim S^{-3/2}$. The existing data demonstrates that this is not the case (see Fig.~\ref{n_s32}). However, this is not in contradiction with cosmological models of FRBs (see a recent discussion in \cite{2018ApJ...858....4N}). In  \cite{2016MNRAS.461..984O}, the authors used $V/V_\mathrm{max}$ test to show that the flux distribution of FRBs could be compatible with a uniform distribution in the Euclidean space. The analysis presented in \cite{2018MNRAS.474.1900M} suggests that the real flux distribution of FRBs is not flatter than $S^{-3/2}$.

\begin{figure}
\center{\includegraphics[width=120mm]{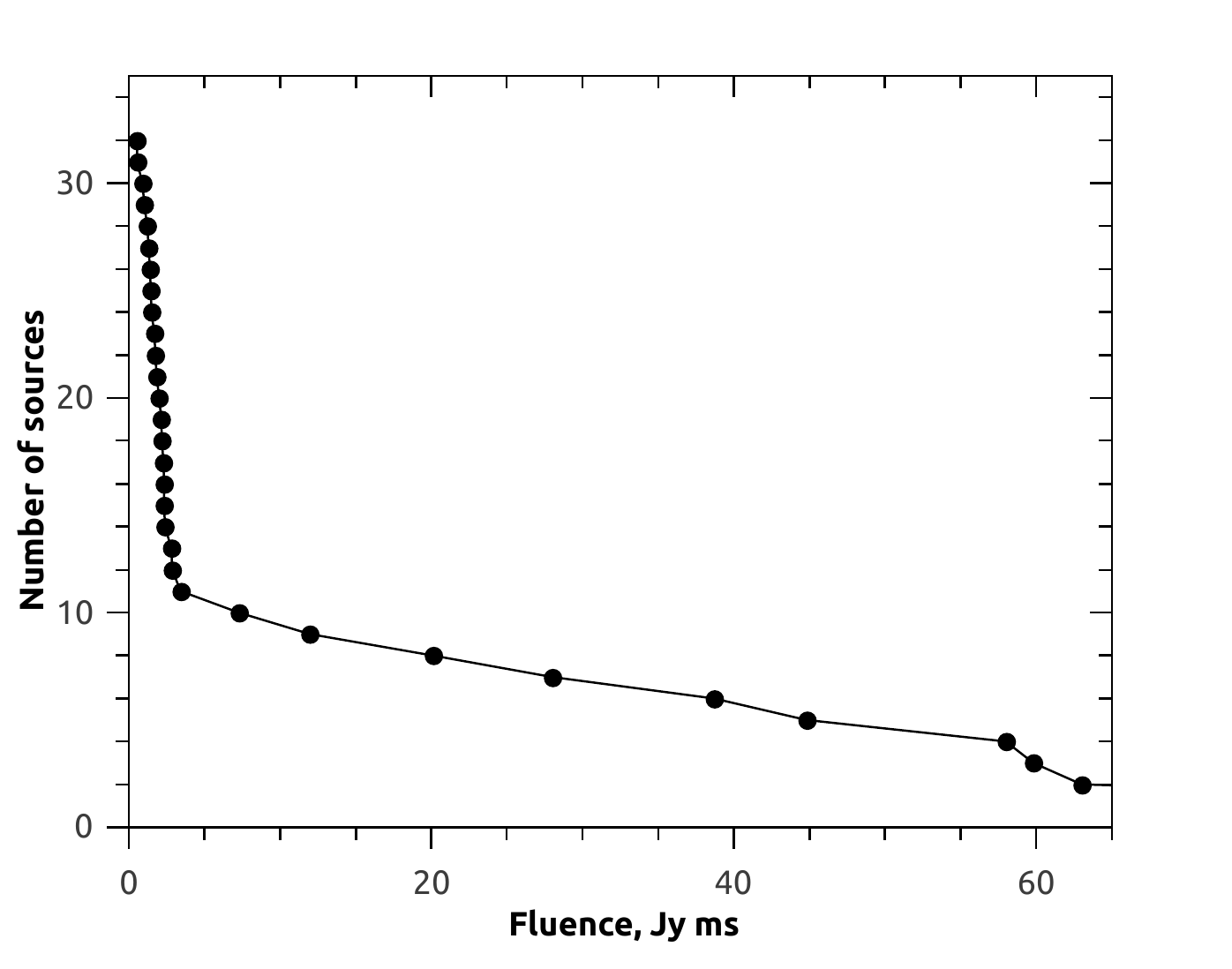}}
\caption{The integral fluence distribution of fast radio bursts:$N(>F)-F$. Note the linear scale in both axes.  Data from the online catalog frbcat.org.
}
\label{n_f}
\end{figure}

The FRB fluence distribution is presented in Fig. \ref{n_f}. Here the fluence is determined as a product of the peak flux and pulse width provided in the catalog. This distribution has a non-trivial shape. Two parts are visible. One corresponds to the range $0.5<(F/\mathrm{Jy} \, \mathrm{ms})<3$, and the second -- to  $3<(F/\mathrm{Jy}\, \mathrm{ms})<100$. It is interesting to note that in each interval, the distribution can be fitted by a linear function $N=a-b\times F$, where $a$ and $b$ are positive coefficients different for two fluence ranges (see also \cite{2016MNRAS.462..941L} where such a property, as far as we know, was noticed for the first time).  
This bimodality, as well as the linear dependences in each fluence intervals, cannot be easily interpreted. Maybe, this is due to selection effects or/and low statistics. Also, on the one hand, among 10-12 sources with the highest fluences,  only about one half was discovered at Parkes. On the other hand, among the rest of known FRBs (which correspond to the second mode with weaker fluences in the plot $N(>F)-F$) events found at Parkes dominate.

\subsection{Repeating Fast Radio Burst source: FRB121102}
\label{s:FRB121102}

The most extraordinary source of bursts known to date is FRB 121102. There are  several  reasons for this. It was the first FRB source detected not at Parkes \cite{2014ApJ...790..101S} -- the detection in the Arecibo data  firmly refuted the scenario of local Parkes artefacts.   More importantly, it is still the only known case of repeating fast radio bursts, which makes it possible to detect new bursts given a sufficient duration of observations.   The dedicated VLBI (Very Long Based Interferometry) observations provided the precise source localization, enabling the subsequent multi-wavelength observational campaign.

The burst was discovered in the archival data of the large pulsar survey (PALFA, 1.4 GHz Pulsar Arecibo L-Alpha Survey) aimed at detecting pulsars and related phenomena, e.g. RRATs, in the Galactic plane ($|b|<5^{\circ}$). The observations were performed with a seven-beam receiver  operating in  322  MHz bandwidth centered at 1375 MHz with a high temporal resolution of 65.5 $\mu$s. The FWHM of individual beam was 3.5 arcmin. Searches for  short transients like RRATs included the dedispersion and scanning in the wide range of the possible DM values from 0 to 2038 $\mathrm{pc~cm}^{-3}$.

A single strong burst with SNR$=14$  was found in the archival record of 02 November, 2012. The burst lasted for $3\pm0.5$~ms,  the measured flux was $0.4^{+0.4}_{-0.1}$~Jy. The burst swept through the frequency band of the detector at a rate corresponding to $DM=557.4\pm2.0~\text{pc~cm}^{-3}$. It was observed in the fourth beam of the array, thus the Galactic coordinates of beam's center $b=-0^{\circ}.223,~l=174^{\circ}.95$  was assigned to the burst. Unfortunately, the localization of the  burst was complicated by its detection in the sidelobe of the beam, and its location error exceeded 5 arcmins. One could expect a rather high value of the dispersion measure at this low Galactic latitude, however the observed DM was almost a factor of three higher than the estimated Galactic contribution $DM_{\mathrm{NE2001}}=188~\text{pc~cm}^{-3}$ (NE2001 refer to the model \cite{2002astro.ph..7156C} which was used for the estimate).  In the absence of compact dense structures which could significantly boost the observed value of the DM, this clearly suggested the extragalactic origin of the burst.  Amount  of the 'excess DM' due to the extragalactic contribution allowed a rough estimate of the distance to the burst: $D\sim1~$Gpc, which eventually almost coincided with the measured distance to the host galaxy (see below). The observations provided only upper limits on the scattering value  $\tau_d<1.5$~ms, much lower than the expectation for a Galactic source with such a DM, also suggesting the extragalactic origin of the burst.

There were no new bursts  detected during the subsequent observations in 2012-2013 with a total duration of several ks. Nevertheless, searches for repeating bursts had not been terminated, and 10 additional burst with the same DM  were finally detected in May-June 2015. Thus, a novel phenomenon, a repeating FRB, was discovered \cite{2016Natur.531..202S}.

Durations of the bursts from the source fell within 2.8-8.7~ms range, typical for  durations of FRBs known to date. On the other hand, their fluxes, 0.02-0.3~Jy, were an order of magnitude lower than those of the typical FRB. The  spectral index $\alpha$   in the power-law approximation $S(\nu)\propto\nu^{\alpha}$ changed  from -10 to 14 for different bursts. No periodicity in the burst times of arrival were found, and the situation has not changed since then despite a large increase in the number of detected bursts. 

A lot of instruments, primarily those operating at radio frequencies,  commenced large observational programs aimed at the source of FRB121102. Almost immediately this resulted in the next large step forward. It is well known that the angular resolution of even the largest single-dish radio telescope is rather mediocre and this considerably hampers accurate identification of sources and their observations in other energy ranges. Interferometric observations at the VLA (Carl G. Jansky Very Large Array) with its superior resolution allowed an improvement of the determination of the source position to 0.1 arcsec \cite{2017Natur.541...58C}. Nine bursts were detected in the 2.5-3.5 GHz band during 83 hours of the VLA observations, three of them with simultaneous observational coverage at the Arecibo, which managed to detect only one of them. This shows that the bursts have a nontrivial spectral shape which, according to the present data, is better described by a Gaussian with $\sim$0.5 GHz width. The high degreе of variability of the complex spectral structure has been recently confirmed by observations with the Green Bank radio telescope (GBT) in the 4-8 GHz band: all 21 bursts that were detected showed a non-trivial structure in the 0.1-1 GHz range, and the spectrum of some of them peaked at frequencies above 6 GHz \cite{2017ATel.10675...1G}.

Also, a weak persistent source (the flux density at 3 GHz  $S_{\text{3~GHz}}=180~\mu$Jy) was detected inside the error box of the VLA observations. More than that, large optical telescopes Gemini and Keck observed a dim object ($r_{AB}=25^m$) whose position coincided with the persistent radio source. The detection  of prominent Balmer and [O III] emission lines  permitted to estimate the redshift of the source, $z=0.192$, which corresponded to a photometric distance of $D$=972~Mpc \cite{2017ApJ...834L...7T}. The object was identified 
as a dwarf galaxy with a diameter of $\sim 4~$kpc and a stellar mass of $M_*=(4-7)\times  10^7~M_{\odot}$. The probability of the chance positional coincidence of the dwarf and the persistent source was estimated to be lower than 
$3\times10^{-4}$. The properties of the galaxy and the localization of the source inside it were refined after deep observations with the \emph{Hubble} and \emph{Spitzer} space telescopes: the source was associated with a compact star-forming region 0.7 kpc in diameter, located in the outskirts of the galaxy at a distance of 2 kpc from the nominal centroid of the diffuse emission, which in turn has a somewhat larger size of $\sim5-7$~kpc than previously estimated from observations with lower resolution  \cite{2017ApJ...843L...8B}. Also, the estimate of the galaxy stellar mass was revised upwards to $M_*=10^8~M_{\odot}$. The metallicity of the host galaxy is low, $\log_{10}[O/H]=-4.0\pm 0.1$. A peculiar class of supernovae,  hydrogen-poor superluminous supernovae, is known to prefer to occur in  this type of galaxies, and this may signal its causal relation with FRB121102.

Further observations at the  EVN (European VLBI Network) and VLBA (Very Long Baseline Array) telescopes revealed
that the size of the persistent source was smaller than 0.2 mas (cor\-res\-ponding to a linear size of 0.7 pc at 1 Gpc distance), its  emission in 1-20 GHz range was   non-thermal and the source's diurnal variability was  at the $\sim10\%$ level. Its distance  to the burst  cannot be larger than 12 mas (40 pc at the 1 Gpc distance) which with a very high degree of certainty suggests their common origin \cite{2017Natur.541...58C,2017ApJ...834L...8M}. The observed flux of the persistent source at the 1 Gpc distance corresponds to a radio luminosity of $L_R\sim10^{39}~$erg~s$^{-1}$, and the luminosity of the brightest bursts can reach $5\times10^{42}~\mathrm{erg~s^{-1}}$. The total energy of individual bursts from the source can be as large as $\mathcal{O}(10^{40})$~ergs.

Deep  observations with {\it Chandra} and {\it XMM-Newton} X-ray telescopes only put upper limits on the luminosity of the persistent source in the X-ray energy range: $L_X<3\times10^{41}$~erg~s$^{-1}$ \cite{2017ApJ...846...80S}. Several models (including FRBs from magnetars) predict that  powerful bursts  with much higher energetics temporally coincident with FRBs can emerge at other frequencies.  The lack of detection  of  gamma- ({\it Fermi} LAT) and X-rays ({\it Chandra})  coinciding with FRBs  constrains the  total energy of the associated flares:  $L_{\gamma}<5\times10^{47}$~erg~s$^{-1}$ and $L_X<4\times10^{45}$~erg~s$^{-1}$, respectively \cite{2017ApJ...846...80S}. Only upper limits were obtained for the possible optical counterparts \cite{2017MNRAS.472.2800H,2018arXiv180600562N}. Final  results of joint observations by the Arecibo radio telescope and VERITAS at energies $>100$~GeV have not been reported as yet  \cite{2017arXiv170804717B}.

The analysis of another 16 bursts that were discovered at Arecibo in the course of regular observations in the 4.1-4.9 GHz band \cite{2018Natur.553..182M} demonstrated a striking property: their linear polarization degree  was close to 100\% after correction for the Faraday rotation. The measured RM values were extremely high $>10^{5}~\mathrm{rad~m^{-2}}$. Also, they were strongly  variable: the results obtained from the observations at GBT showed that the RM valued decreased by 10\% in only 7 months. These observational properties could be explained in a model where the source is located in the immediate vicinity of a black hole inside the accretion flow. Another viable possibility is that the source is surrounded by a young pulsar nebula with a pronounced filamentary structure.

A lot of theoretical studies try to build   viable models of FRB 121102 which could explain the observational properties of the bursts. The recurring bursting activity naturally allowed to exclude all catastrophic scenarios, e.g. FRB from NS mergers, and a large number of 'non-traditional' ones, such as collisions with asteroids and interaction  with axion miniclusters.   Localization of the source in the outskirts of the host galaxy could be perceived as an evidence against the models where the generation of bursts is linked to the activity of central AGN. Successful models  must also simultaneously explain the observed properties of the persistent source, lack of the evolution of DM and strong evolution of RM, and, finally, statistical distributions of burst energetics, fluence, and duration.  
The most promising models (as in the case of non-repeating FRBs) are those related to supergiant pulses of radio pulsars and strong flares of magnetars. As in the models of  single flares, the repeating bursts are powered by the rotational energy in the former case and by the energy of the magnetic field  (i.e. energy of the electric currents, flowing inside the star) in the latter case. The magnetar model is slightly more preferable due to  energetic considerations \cite{2017ApJ...843L..26B, 2017ApJ...841...14M}. Indeed, a very high  conversion efficiency of the rotational energy into radio, which, e.g. by several orders of magnitude exceeds this coefficient for the Crab pulsar,  is needed  in the model of giant pulses \cite{2017ApJ...839L...3K}. In both cases, the bursts come from  a very young neutron star and the persistent source is  a pulsar (magnetar) nebula and/or young supernova remnant.

The analysis  of burst times of arrival can also produce valuable insights. The distribution shows a strong clustering of bursts, and the underlying random process considerably deviates from a stationary Poissonian process but appears to be quite close to the similar distribution of flares produced by soft gamma-ray repeaters (SGRs) \cite{2017JCAP...03..023W}. As mentioned above, no periodicity was  found after the analysis of several hundreds of bursts. The strongest indirect indication of its existence comes from the  observation of  two pairs of bursts with a very small separation (34 and 37 ms), which can be a consequence of a fast rotation of the neutron star with a period $P<3~$ms \cite{2018MNRAS.476.1849K}.

It is still possible that most of the rich phenomenology of FRB121102 is generated in the interstellar medium of the host galaxy rather than in the source itself -- strong enhancements of electron density could play the role of plasma lenses leading to considerable amplification of signal and its pronounced spectral modulation \cite{2017ApJ...842...35C}. A similar effect was proposed to  explain strong spectral modulations of recently detected FRB170827 \cite{2018arXiv180305697F}. Also extreme lensing events by plasma density fluctuations with magnification values reaching 70-80 and spectral features resembling ones in the spectrum of the FRB121102 were recently observed in the eclipsing binary PSR B1957+20 ('Black Widow')\cite{2018arXiv180509348M} giving additional support to this model.

Whether the FRB121102 is unique or belongs to a large class of bursts is another very important question. In the analysis of FRB observations, a simple assumption of uniform population, i.e. that  all bursts are repeating ones, is frequently used. However, it was shown that this assumption is most likely flawed, otherwise several repeating bursts should already have been detected in the  Parkes data \cite{2018ApJ...854L..12P}. Given that the Parkes telescope observed a large fraction of the celestial sphere, this suggests quite opposite:  the repeating bursts is quite a rare (or short-living) phenomenon.

\section{Hypotheses of the FRB origin}

The task to explain the nature of FRBs can be divided into two parts. The first one can be called a {\it physical} part. It is related to the emission mechanism which results in a high brightness temperature and explains other spectral and timing parameters of FRBs. The second one is an {\it astrophysical} part relating to astronomical objects and phenomena which have features appropriate for the FRB sources: spatial distribution, event rate, energy release, transparency of the medium for emission, etc. Clearly, any realistic model should take into account both aspects, and this appears to be very non-trivial.

The list of hypotheses proposed as an explanation of FRBs counts more than twenty items, even without detalization. Several of these ideas are mentioned in the brief review \cite{2016MPLA...3130013K}.
Some of the proposed models are based on exotic physics (cosmic strings, white holes, charged black holes, etc.), some -- on more conventional scenarios. We start with a list of models which presently  are not considered as viable explanations of the whole class of FRBs. Separately we will discuss coalescing neutron stars because several years ago this scenario was considered to be one of the most likely FRB mechanisms. Then, we present in more details two models (magnetar bursts and supergiant pulses of energetic radio pulsars) which are accepted now as the most promising candidates. Note that the FRB population can be non-uniform, i.e. not necessarily all sources should be described by a single model. In this respect, the repeating source is of a special interest (see above, Sec. \ref{s:FRB121102}). After discussing different theoretical scenarios, we demonstrate how FRBs can be used in cosmology and extragalactic astronomy, as well as probes of fundamental physical theories.

\subsection{Less probable hypotheses}
 
 Hypotheses about the FRB nature can be divided into two broad categories: related to neutron stars and exotics. On the one hand, neutron stars are well-studied sources of strong radio emission, including bursts, on the other hand -- these objects can easily provide necessary energy release and explain the short duration of the flares. The latter is due to the fact that characteristic timescales: dynamical (close to the surface), and Alfvenic (in the inner magnetosphere),  are $\lesssim 1$ ms. The appearance of exotic models can be explained by the possibility to apply interesting non-standard ideas to real observations.

Very soon after the announcement of the first (Lorimer) burst, a model of generation of FRBs by cosmic strings was proposed \cite{2008PhRvL.101n1301V}. In this model,  cusps  of cosmic strings can be  sources of electromagnetic emission.
Later this approach was further developed in several other studies by different authors. 

Another exotic scenario was mentioned in \cite{2012MNRAS.425L..71K}, in which the discovery of the second FRB was reported. The model involved a primordial black hole evaporation. 
Bursts of electro-magnetic emission emerging due to this process at different wavelengths were predicted already 40 years ago in \cite{1977Natur.266..333R}. However, to explain observed FRB fluxes it is necessary to place the sources at small distances, $<300$~pc.

An interesting modification of the evaporating black hole scenario was studied in \cite{2014PhRvD..90l7503B}.  Here the authors considered the situation when a white hole appears at late stages of a black hole evaporation -- this process is possible in the loop quantum gravity theory. In this case, the remaining mass (at the moment of a burst) of the objects is higher, so this results in a more powerful signal which can be seen from larger distances than in the usual Hawking process.

The last model related to single black holes, which we want to mention, was presented in \cite{2016ApJ...826...82L}. 
In this paper, the authors considered the collapse of the magnetosphere of a charged rotating (i.e., Kerr-Newman) black hole. We stress that in the framework of this scenario, as well as in most of other exotic approaches,
under realistic conditions it is difficult to explain the main characteristics of FRBs.

On the other side (with respect to scenarios including exotic physics) we find a model of flares from normal stars in our Galaxy \cite{2014MNRAS.439L..46L}. This is an interesting example of the FRB scenario that was rapidly criticized from different points of view and rejected (see  \cite{2014MNRAS.441L..26T, 2016ApJ...818...74L} and references therein). It is worth mentioning here that presently the overwhelming majority of scenarios is based on the extragalactic nature of FRBs.

Magneto-hydrodynamical mechanisms of FRBs were proposed in several studies. In one of them the authors considered the bremsstrahlung  radiation of collisionless plasma \cite{2016PhRvD..93b3001R}. According to this model, FRB can be a feature of a relativistic jet -- a beam of relativistic electrons produces coherent radio emission due to interaction with plasma turbulence.

In \cite{2014A&A...569A..86M}, a binary system with a radio pulsar was analyzed. If an object (a planet, an asteroid, or a white dwarf) is embedded in a magnetized pulsar wind, two stationary Alfv\'en waves (``Alfv\'en wings'') might appear. 
Due to instabilities, these stationary waves can become sources of radio emission. Because of relativistic beaming, a strong peak of emission can be observed only if an observer is located exactly on the line pulsar-object. It is very difficult to explain many properties of FRBs by this model, especially the absence of periodicity related to the orbital motion of the body around the neutron star.

We continue with non-standard scenarios involving neutron stars.
In some of them, a neutron star belongs to  a binary system, in some -- not. 
For example, in the model proposed in \cite{2009AstL...35..241E}, an FRB appears after a supernova explosion in a binary system where the second companion is a neutron star with large magnetosphere. 
The supernova shock wave interacts with the magnetosphere forming  a magneto-tail. Plasma instabilities in the tail can result in coherent radio emission. 
A similar model, called the ``cosmic comb'', was later developed in \cite{2017ApJ...836L..32Z}. A problem of this model (as well as many others) is related to the impossibility to explain high event rate (in particular, here it is necessary to have some specific orientation of the binary relative to an observer), and, of course, it is impossible to explain the repeating source.

As mentioned above, neutron stars attract interest from the point of view of FRB scenarios because of their strong magnetic fields enabling a short radio pulse generation. This feature is employed in the following two models related to isolated neutron stars. In the first model \cite{2015ApJ...809...24G}, an asteroid is falling onto a neutron star. As a result, a cloud of ionized matter is formed and expands mostly along the magnetic field lines. The authors suggest that the coherent emission of a narrow layer at the surface of this fireball can be responsible for the FRB phenomenon.

According to the second model, the crucial ingredient is the Primakoff process of conversion of an axion into a photon (or vice versa) in a magnetic field. Axions are one of the most popular particle candidates to explain dark matter. Thus, they can be very abundant in the Universe. They can form clouds with masses  $10^{-12}$~--~$10^{-11}\, M_\odot$ which can eventually fly  through a magnetosphere of a  neutron star. Then, via the Primakoff process, part of axions is converted to photons producing a burst of electromagnetic radiation (see \cite{2009JETP..108..384P} for an early discussion of non-transient emission due to axions conversion in a neutron star magnetosphere). This was proposed as an explanation for FRBs in papers  \cite{2015JETPL.101....1T}  and \cite{2015PhRvD..91b3008I}. However, in \cite{2017IJMPD..2650068P}  it was demonstrated that such an axion cluster might be destroyed by tidal forces before  the burst maximum and  the duration of the transient will occur on second rather than   millisecond time scale.

FRBs are often attributed to some catastrophic process in neutron stars. Among them is the deconfinement of the neutron star matter \cite{2015PhRvC..92d5801D}. The deconfinement results in dramatic changes in the neutron star interiors. In a very short time (of the order of a millisecond), the radius and gravitational mass of the object are changed. This is accompanied by a  huge energy release (mostly, in the form of neutrinos). However, an electro-magnetic counterpart can also  appear. Taking into account that the magnetic field of the object might be also significantly modified, one can speculate about the possibility to have a short radio flare. Two scenarios of a FRB accompanying the deconfinement of a neutron star have been proposed \cite{2015arXiv150608645A, 2016RAA....16e..11S}.

Finally, it is necessary to mention the scenario with a supramassive neutron star \cite{2014A&A...562A.137F}. 
A supramassive neutron star is a compact object which avoids the gravitational collapse thanks to its rapid rotation. During the spin-down phase, the central density increases, and when some critical density is reached, the object collapses to a black hole. At this moment a short radio burst can be generated. Note that in this model, it is potentially possible to have a large time interval between the neutron star birth and the FRB emission. This is important since all FRBs discovered in the real time were not accompanied by supernova, gamma-ray bursts, etc. In \cite{2014A&A...562A.137F}, the authors assumed that the collapse can happen hundreds, or even thousands, years after the neutron star formation. However, the analysis presented in \cite{2014MNRAS.441.2433R} demonstrated that if a neutron star is formed during a binary system coalescence, the collapse most probably happens within a few hours. The reason for such a short time interval is related to a very rapid spin-down due to strong magnetic field which inevitably must be formed in this situation  \cite{2014MNRAS.441.2433R}.
A model of FRBs accompanying gamma-ray bursts after the coalescence of binary neutron stars has also been considered in \cite{2014ApJ...780L..21Z}.

 \subsection{Coalescence of magnetized compact stars}

In the end of the previous section, we already mentioned models related to coalescence of binary neutron stars.
Presently, such scenarios are considered as plausible ones. However, it is clear that they can not be responsible for the whole population of FRBs, and definitely, the repeating source, FRB121102, must be explained in a  different way. 

  For some time, the binary coalescence model in which at least one compact object is a neutron star was very popular, because the  proposed identification of a long radio transient with  FRB 150418 pointed towards an elliptical galaxy as the possible host galaxy of this event \cite{2016Natur.530..453K}.  However, further studies \cite{2016A&A...593L..16G, 2017MNRAS.465.2143J} demonstrated that these two radio transients were not related with each other, and the long one was due to an activity of the galactic nuclei. Thus, the association of FRB 150418 with an elliptical galaxy was found to be erroneous. 

Initially, the model of the neutron stars coalescence was mentioned in relation to FRBs in \cite{2010Ap&SS.330...13P,2013ApJ...768...63L}. Later, it was developed and discussed in more details in \cite{2013PASJ...65L..12T}. It is necessary to note that in the early studies different authors (see, e.g. \cite{2010Ap&SS.330...13P} and references in \cite{2013PASJ...65L..12T}) studied the generation of radio emission accompanying the neutron stars coalescence, however, strong millisecond radio bursts have not been discussed.

The binary NS coalescence is assumed to give rise to a massive rapidly rotating magnetized object which then collapses into a black hole. This allows high luminosity and short duration of the burst to be explained simultaneously.  
A maximum energy release can be estimated from the standard equation for the magneto-dipole losses:

\begin{equation}
\dot E\approx 4\times 10^{45} (B/10^{13}\mbox{G})^2
(R_\mathrm{NS}/10^6\, \mbox{cm})^6 (P/1\, \mbox{ms})^{-4} \mbox{erg} \, \mbox{s}^{-1}.
\end{equation}
Here $B$ is the surface magnetic field, $R_\mathrm{NS}$ is the neutron star radius and $P$ is its spin period.
As in the case of radio pulsars, only a tiny fraction of the total energy losses is emitted in radio waves. Nevertheless, the observed FRB fluxes $\sim 10^{-13}$~erg cm$^{-2}$s$^{-1}$, which corresponds to luminosity of the order of $10^{40}$ erg s$^{-1}$ for distances about several hundred Mpc,  can be easily explained even for low transformation coefficients.

In \cite{2013PASJ...65L..12T}, the emission mechanism were not studied in details. Later, in paper \cite{2016ApJ...822L...7W}  the authors analyzed the unipolar inductor model in the application to FRBs in the scenario with neutron star coalescence (earlier, this approach was used in \cite{1996A&A...312..937L}, where it was demonstrated that the radio luminosity exhibits a power-law growth for different emission mechanisms with a typical timescale of the main burst of about a fraction of a second).

Note that the estimated FRB rate (about 100 events per day in the volume $\sim 1$ Gpc$^3$) is too large to be explained by neutron star coalescence. The conservative number for the latter one is about once in $10^5$ years in a Milky way-like galaxy  \cite{2014LRR....17....3P}, which can be re-calculated to $\sim 100$ coalescence per year in $\sim 1$ Gpc$^3$, which is much smaller than the rate of FRBs. Most of up-to-date estimates of the binary NS coalescence rate, improved upon after the first detection of the gravitational wave signal from the binary neutron star merger event GW170817 \cite{2017PhRvL.119p1101A}, is an order of magnitude higher, $\sim 1540$ per year from Gpc$^3$, but this does not help much. No short radio transients have been detected from GW170817 \cite{2017ApJ...848L..12A}, however, the first radio observations started a few hours after the registration of the gravitational wave signal \cite{2017ApJ...848L..21A}. Also, it cannot be excluded that the non-thermal radio emission after a neutron star coalescence is strongly beamed.

Another variant of the FRB model based on the coalescence of compact objects was proposed in \cite{2015ApJ...814L..20M}. In this paper, the authors considered the merger of a neutron star with a black hole. It is assumed that the neutron star is not tidally destroyed until the latest stages of the coalescence. Thus, a black hole can enter into the neutron star magnetosphere (inside the light cylinder, $R_\mathrm{l}=c/\omega$). In the case of a non-rotating black hole, the radio luminosity can be calculated as \cite{2015ApJ...814L..20M}:
\begin{equation}
L=1.3\times10^{40}[\hbox{erg\,s}^{-1}] \left( 1-2M_\mathrm{BH}/r \right) (v/c)^2 (B/10^{12} \mbox{G})^2 \eta_{-2} (M_\mathrm{BH}/10\, M_\odot)^2 (R_\mathrm{NS}/10^6 \mbox{cm})^6 (r/30\, M_\odot)^{-6}. 
\end{equation}
Here $M_\mathrm{BH}$ is the black hole mass, $B$ is the neutron star surface magnetic field, $R_\mathrm{NS}$ is its radius,  $r$ is the distance from the black hole to the neutron star surface given in solar masses. The efficiency of the radio emission $\eta$ is normalized to 0.01 but can be higher for rotating black holes.

Presently, it is accepted that the rate of neutron star -- black hole coalescence is smaller than that of double neutron stars \cite{2014LRR....17....3P}. Therefore, the authors of \cite{2015ApJ...814L..20M}  suggested that only a small subpopulation of FRBs can be described by this model. Additionally, the authors mentioned that in this mechanism bimodal bursts, like FRB 121002 \cite{2016MNRAS.460L..30C}, can appear.

Finally, coalescences of magnetized white dwarfs have also been  proposed as a model for FRBs \cite{2013ApJ...776L..39K}. 
In this case, the rotation energy loss rate is too low   to explain the power of the transients, and the authors additionally considered the energy release due to the reconnection of magnetic field lines. In this model, the coherent radio emission originates in the polar region of a rapidly rotating (the spin period $\sim 1$ s) white dwarf with a very high magnetic field of $>10^9$~G. To provide the necessary power, about 1\% of the magnetic field energy must be converted into the radio emission. In this scenario, some FRBs can be accompanied by a SN Ia if the total mass of the coalescing double white dwarfs exceeds the critical one (close to the Chandrasekhar limit $\simeq 1.4 M_\odot$). Up to now, no such FRB counterparts have been observed.

Many models considered as not very realistic are related to more plausible scenarios. In the next section we will discuss some of them, as well as their modifications.

\subsection{The most promising models}
\label{s:models}
As we already mentioned in the Introduction, any realistic model of the FRB must simultaneously explain the characteristic observational properties of the bursts, such as a very high brightness temperature of the bursts, their energetics, the occurrence rate, the spatial distribution,  etc. (see  above in Section  \ref{s:Observations}).

\subsubsection{Model of supergiant pulses}
\label{s:GP_model}

Several radio pulsars are known to  occasionally emit giant pulses (GP). The best studied example of this kind is a young pulsar in the Crab nebula. The giant pulses are short, $\sim \mu$s radio bursts with the peak flux that exceeds that of typical pulses by several orders of magnitude. Dedicated observations revealed a complex nanostructure of  Crab's bursts: the giant pulses were further resolved into a sequence of much shorter nanoshots which collectively form the giant pulse.

These nanoshots were observed at 9 GHz frequency in a 2 GHz bandwidth. Nanoshots remained unresolved despite the large bandwidth and correspondingly the high temporal resolution implying that their durations $W$ were less than 0.4~ns; the highest observed  peak flux density exceeded 2~MJy \cite{2007ApJ...670..693H}. That means that in the non-relativistic case the emission should take place in a region with a size of less than $cW\sim10$~cm and the brightness temperature should be higher than  $2\times10^{41}$~K. Correction for relativistic effects does not considerably alter these extreme properties: if the Lorentz factor of the bulk motion is as high as $\gamma\sim10^2$--$10^{3}$, the size increases up to $10^3$--$10^5$~cm,   while the brightness temperature simultaneously decreases down to still huge values $10^{35}$--$10^{37}$~K. Thus, we can  say with a high degree of  confidence   that nanoshots are generated coherently in rather small regions.

The problem of GP generation, as well as more general problem of the pulsar radio emission, has not been conclusively solved as yet. Still, their observed  properties were used to suggest that the FRBs are  GPs from very large distances \cite{2015ApJ...807..179P, 2016MNRAS.457..232C,2016MNRAS.458L..19C}.\footnote{See also the short comment in  \cite{2010vaoa.conf..129P},  where it was suggested the the linear scaling of the Crab GPs with energy spin-down rate can explain the FRB properties in case where the young pulsar has a large, but still realistic spin-down rate.} In this model, an FRB  observed at  $\sim{100~\text{Mpc}}$ distance is a collection of a very large number $N_i \sim 10^{10}D_{\text{100~Mpc}}^2$ of nanoshots,  which are similar to the strongest nanoshots observed from the Crab pulsar (see above).

FRBs in this model can  be produced only if the relativistic plasma inside the pulsar magnetosphere is endowed with fairly extreme properties, which, nevertheless, can be realized in nature. Estimates of the total power of the process show that the most natural candidates could be neutron stars with ages smaller than 100 years and spin periods shorter than 20 ms, giving the model its name, the 'young neutron star model'. In this model, the bursts occur at  distances about 100-200~Mpc, rather than $\mathcal{O}$(Gpc). In order to reproduce the observed rate of FRBs, each individual source must produce  $10^{4}-10^{5}$ bursts during its active phase. This number is low enough not to contradict the fact that there is only one repeating FRB known, but   FRB 121102 can belong to another class of the sources (see Section \ref{s:FRB121102}). A search for nanostructure in the bursts  could serve as a direct observational test of the model, however this approach will be severely hindered by the effects of propagation of the radio emission in the interstellar and intergalactic medium.

The model was further developed in \cite{2016MNRAS.462..941L}: the bursts were generated by very young pulsars with ms periods, which were 
surrounded by a supernova remnant. In this model, consecutive bursts would have a decreasing DM evolving on a $\mathcal{O}$(yr) timescale. The model provides a natural explanation for the lack of detections at low frequencies (less than  $\sim$600 МHz): the high density of a young SNR results in the effective free-free absorption at these frequencies \cite{2016MNRAS.462..941L,2016ApJ...824L..32P}.

A relatively small distance to the source in this model can also be used for additional observational tests \cite{2016MNRAS.462L..16P}. First, a significant degree of correlation of the FRB positions  with nearby galaxies is expected. Presently, the poor FRB localization and their low number do not allow this type of checking to be performed, but estimates show that this test can give meaningful results with $\mathcal{O}$(100) detected bursts. Second, young neutron stars are known to be strong X-ray sources, in this case, we could expect the FRB sources to appear as ultra-luminous X-ray sources (ULX). To verify this hypothesis, a better FRB localization is needed as well.

\subsubsection{Magnetars}
\label{s:magnetar_model}
Shortly after the discovery of the first FRB in 2007, the authors of 
\cite{2010vaoa.conf..129P} (see also \cite{2013arXiv1307.4924P}) put forward the hypothesis that FRBs can be related to hyperflares of magnetars. These hyperflares are powerful short (possibly repeating on timescales of the order of decades or even hundred years) episodes of electromagnetic energy release 
($\Delta E\sim 10^{44}-10^{46}$~ergs) from the most strongly magnetized neutron stars -- magnetars. The magnetar hyperflares are likely to be related to a catastrophic evolution of the superstrong neutron star magnetic field (up to 
$B\sim 10^{15}$~G at the surface) in the neutron star magnetosphere.
(see e.g. \cite{2015RPPh...78k6901T} for a review).
In the first place, the arguments of the authors 
\cite{2010vaoa.conf..129P} were based on the estimated  statistics of the magnetar hyperflares $\sim 100$ from the local 1 Gpc$^3$ volume every day, which is comparable to the FRB statistics. In addition, the temporal characteristics (the sharp rise of the flare profile) are similar in both phenomena.  
A total energy release in a magnetar hyperflare is also able to readily explain the observed FRB radio fluxes. A simple scaling of the radio luminosity in one of the models 
 \cite{2002ApJ...580L..65L} proposed to explain weak magnetar flares leads to good correspondence between the observed FRB fluxes by assuming that the dispersion measure is mainly due to the signal propagation in the intergalactic medium. 
Besides, the magnetar hyperflare model easily explains the lack of FRB detections in other bands from distances $\sim$1 Gpc.

Papers \cite{2010vaoa.conf..129P, 2013arXiv1307.4924P} did not propose a physical model of the radio emission generation. In the framework of the magnetar hyperflare scenario, such a model was elaborated in paper 
\cite{2014MNRAS.442L...9L}. Later on, a similar approach was considered in other papers, in particular, in \cite{2016MNRAS.461.1498M}. In this model, a radio flare is generated by the synchrotron maser mechanism\footnote{
The generation conditions for GHz maser radiation in a magnetized relativistic plasma were considered in paper \cite{2017MNRAS.465L..30G}.} in a relativistic shock arising in a magnetized plasma around a magnetar. For example, it can be an analog of a pulsar wind nebula around the magnetar. Similar nebulae indeed are observed around some magnetars and highly magnetized pulsars, see for example
\cite{2009ApJ...707L.148V, 2012ApJ...757...39Y, 2016ApJ...824..138Y, 2013IAUS..291..251S}.

In this model, an electromagnetic pulse triggered by a powerful magnetar flare reaches the boundary of the pulsar wind nebula located at a distance of $r\sim 10^{15}-10^{16}$ cm.
Interaction between the electro-magnetic pulse and the nebula  gives rise to two shocks (forward and reverse)  expanding from the contact discontinuity, and the contact discontinuity itself moves with 
a Lorentz-factor of  $\gamma_{cd}\sim 10^4$. Behind the front of both -- forward and reverse, -- (relativistic) shocks propagating in the magnetized plasma of the nebula, an inverse population of electrons with energies less than   $\gamma_{cd}  m_ec^2$ appears, 
and the conditions for the synchrotron maser radiation from the particles at a frequency of about 1 GHz are satisfied. This frequency is determined by the relativistic cyclotron frequency in the magnetic field behind the shock front  
 $\sim eB_{EMp}/(\gamma_{cd}m_ec)$ (here $B_{EMp}$ is the magnetic field in the electromagnetic pulse in the interaction region, $B_{EMp}\sim 10^5$~G). 
The duration of the radio pulse  $W$ is determined by the time it takes for the electro-magnetic pulse energy to be transferred to the pulsar nebula plasma that turns out to be of the order of the initial electromagnetic pulse width $l/c\sim 10^{-4}$ s.

An important prediction of the model 
\cite{2014MNRAS.442L...9L} is the appearance of a simultaneous ms hard radiation pulse from the FRB at TeV energies due to the synchrotron radiation of relativistic particles behind the forward shock which falls into the TeV range in the observer frame. Potentially, ground-based observations with gamma-ray telescopes (H.E.S.S., MAGIC, etc.) can be used to test this prediction. A search for steady TeV emission from the repetitive flaring FRB 121102 (which, apparently, is a special case) carried out by the VERITAS Cherenkov telescope array gave null result 
 \cite{2017arXiv170804717B}; the results of synchronous searches for pulsing TeV and radio flares from this source have not been published so far. 

Unlike the GP model, in the case of the synchrotron maser radiation, the time profile of the radio pulse generally should repeat the form of the initial electromagnetic pulse generated during the magnetosphere restructuring during a giant magnetar flare and should not demonstrate nanosecond substructures. This also can be tested by radio observations with high time resolution. 

It is possible that the short radio flares reported in paper 
\cite{2013MNRAS.428.2857R} from the Andromeda nebula (M31) direction were weaker versions of FRBs. Observations with the WSRT (Westerbork Synthesis Radio Telescope) at the frequency 328 MHz detected millisecond radio flares with dispersion measure expected from the distance to M31. In total, several dozen flares were registered. In one case, six flares from one source with a dispersion measure of 
54.7 pc cm$^{-3}$ were detected in one hour. Also, several other repeating flare candidates were found. In the magnetar model, less powerful radio flares can accompany quasi-regular activity of magnetars which are the sources of repeating soft gamma-ray bursts, i.e. SGRs,  that can be observed from nearby galaxies \cite{2013arXiv1307.4924P}. 

The magnetar model was criticized in paper \cite{2016ApJ...827...59T} 
in which the authors searched for radio emission from a hyperflare from the Galactic magnetar SGR 1806-20. As the source is located at least by four orders of magnitude closer than the potential FRB sources, one could expect to observe a very powerful radio flare that could be detected by sidelobes of several radio telescopes. The obtained null result reliably suggests that the hyperflare from SGR 1806-20 observed in December 2004 was not accompanied by a radio flare within a wide opening angle (in the magnetar model, FRBs does not emitted into a narrow radio beam). This, however, is not a crucial argument because the conclusion is based on only one source whose properties can be different. For example, the magnetar SGR 1806-20 is not surrounded by a pulsar wind nebula. 
Another model based on the synchrotron maser mechanism has been considered in \cite{2018arXiv180602700L}. The authors conclude that weakly magnetized accreting neutron stars are more favored as FRB progenitors. Clearly, further multi-wavelength observations are needed to distinguish between different possibilities.

\subsection{FRBs as probes of the intergalactic medium and an instrument to probe fundamental theories and cosmological models}

FRBs are interesting not only by themselves, but also as a tool for astrophysical and physical studies. In the first place, very short and intense radio bursts are ideal probes for the intergalactic medium. Then, strong bursts at large distances can be interesting for cosmological studies. Beside, short pulses which propagate in space for billion years offer an opportunity to test predictions of fundamental theories. In this subsection we discuss these possibilities.

Applications of FRBs for studies of the intergalactic medium have been analyzed in many papers. Usually, in such studies it is assumed that mostly the dispersion measure (and may be, also, the rotation measure) is due to the intergalactic medium, and not to the medium in the host galaxy or to the matter just in the vicinity of the source. However, even if the contribution from the intergalactic medium is not dominating, but  can be somehow figured out, FRBs still can be good probes for the gas and magnetic fields in filaments, voids, and galaxy clusters.

In \cite{2016ApJ...824..105A}, the authors used the results of numerical modeling of the large scale structure up to $z=5$ to estimate the expected values of  dispersion measure and rotation measure, and relative contribution due to different elements of the structure (filaments, voids, clusters). It was shown that in the range $0.1\lesssim z \lesssim 1.5$  the main contribution to the dispersion measure is due to filaments, and on larger distances voids start to dominate. 
The rotation measure is mainly due to galaxy clusters. However, in directions where the line of sight is not crossing clusters, it is possible to use FRBs to measure magnetic field in the filaments.

In \cite{2017ApJ...834...13F}, it was studied how the data on dispersion measure of FRBs observed behind galaxy clusters together with the data on Sunyayev-Zeldovich effect for these clusters can be used to determine parameters of the warm intergalactic medium. Detection of several FRBs behind a cluster might enable the determination of the density profile, and then Sunyayev-Zeldovich effect provides the pressure profile. Together, these data would give the temperature profile inside the cluster.

With increased FRB statistics, it will be possible to employ them for studies of the intergalactic medium even without precise individual  distance measurements. Such an approach is discussed in \cite{2017PhRvD..95h3012S}. However, it is necessary to note that most probably to finally reach this goal, observations from such instruments as UTMOST, CHIME, HIRAX, and FAST could not be sufficient. Only  the  Square Kilometer Array (SKA) can provide the necessary number of FRBs -- about $ 10000$.

Also, about ten thousand FRBs are necessary to put strong constraints on massive compact objects in the Galactic halo (MACHO), which potentially can contribute to cold dark matter, via gravitational lensing of these radio transients \cite{2016PhRvL.117i1301M}. Such an approach can probe an interesting mass range of MACHOs: 20-100 solar masses. For this range, the  present-day constraints are not strong enough to exclude completely a significant contribution of such objects to dark matter. The time delay during lensing on such objects can be about several milliseconds. Potentially, even smaller delays related to lighter lenses can influence FRBs timing. For example, recently the so-called 'nanolensing' of FRBs  has been studied in \cite{2017ApJ...850..159E}.
If the necessary precision can be reached in observations, it is possible to derive a very strong limit on the amount of light compact lenses contributing to dark matter.

Almost immediately after the publication of the paper by Thornton et al. \cite{2013Sci...341...53T}, different authors started to analyze whether the FRB statistics is compatible assumption of cosmological distances to these sources (see references e.g. in \cite{2018ApJ...858....4N}). Simultaneously, a discussion of the possible application of FRBs as cosmological probes was initiated. Among the statistical studies, one can mention papers \cite{2015MNRAS.451.4277D, 2016MNRAS.458..708C}. In the first of them, the population of FRBs was studied using the numerical  modeling of galaxy distribution and under the assumption that all FRBs are standard candles. The authors analyzed three variants of the model. In the first variant, it was assumed that the number of FRBs correlated with the total stellar mass (which, for example, corresponds to the neutron star coalescence model). The second variant assumed that FRBs correlated with the star formation rate (in correspondence with models where the FRB sources are young neutron stars, for example, magnetars). Finally, in the third variant it is supposed that the FRB phenomenon was somehow linked to the existence of the central supermassive black hole. With low statistics it is impossible to distinguish between these three variants. In \cite{2016MNRAS.458..708C}, properties of the FRB population were also studied assuming their extragalactic origin. In particular, it was shown that only with several dozens of sources it could be possible to make conclusions about 
the basic properties of the population. Thus, this could not be done with the number of sources available at the time of writing of \cite{2016MNRAS.458..708C} (and, also, presently).

Among papers of the second kind devoted to discussions of possible applications of FRBs to cosmological studies, we can mention, for example, \cite{2014PhRvD..89j7303Z, 2014ApJ...783L..35D, 2014ApJ...788..189G}. General conclusion is that FRBs (if they indeed are sources at $z\sim 1$) can be used to derive estimates of global cosmological parameters only with significantly improved statistics and with at least the basic understanding of their properties (e.g., the luminosity distribution, birth rate, etc.; see recent modeling and discussion in \cite{2018ApJ...858....4N}). A recent analysis of prospects of cosmological applications of FRBs can be found in \cite{2017arXiv171111277W}. The results of this study are more pessimistic than the previous ones. Nevertheless, the authors conclude that in the future FRBs can be employed to constrain the baryon density of the Universe.

Simultaneous measurements of DM and the redshift $z$ are available for many FRBs could make them a very effective tool for cosmological studies. This opportunity was studied in \cite{2014ApJ...783L..35D, 2014ApJ...788..189G}, where the authors speculated that redshifts can be determined thanks to observations of gamma-ray bursts associated with FRBs. Among other results this would allow one to estimate the baryonic contribution to the total density, and in the case of distant sources -- to probe parameters of re-ionization. Several tens of FRBs with known redshifts (if they are distant enough) could be used to put important constraints on cosmological models. 
We can also mention paper \cite{2017A&A...606A...3Y}, in which the possibility to measure the proper distances to the sources using a large number ($\sim 500$) of FRBs with known redshifts was analyzed.

Observations of FRBs can be used to determine fundamental physical parameters and to make independent tests of physical principles. In the first place, one can mention constraints on the photon mass and tests of the equivalence principle and Lorentz invariance.

As FRBs demonstrate very narrow pulses, it is possible to measure with high precision delays between signals at different frequencies. This gives an opportunity to use the sources for tests of the equivalence principle. In \cite{2015PhRvL.115z1101W}, the authors discussed such an approach by considering the signal propagation in the Galactic gravitational potential. The use of radio data faces the problem of the signal widening due to the propagation in the interstellar and intergalactic medium, and this effect cannot be easily separated from the hypothetical time delay due to the equivalence principle violation. Thus, it is important to have simultaneous observations at other wavelengths or/and to obtain an independent estimate of the dispersion measure. Besides the propagation in the Galactic potential, it is also possible to discuss signals observed behind massive galactic clusters \cite{2016arXiv160104558Z}.  Modern constraints on the parameter $\Delta \gamma$ (difference between the post-Newtonian parameter $\gamma$ for two frequencies) obtained with FRBs are about $\mathcal{O}(10^{-8})$. However, they can be improved by an order of magnitude with more precise measurements \cite{2016ApJ...820L..31T}.

The first constraints on the photon mass from observations of FRBs appeared after the claim that the host galaxy of FRB150418 was identified \cite{2016ApJ...822L..15W, 2016PhLB..757..548B}. As later on the identification turned out to be erroneous, these papers are interesting only from the point of view of methodology. Only when the repeating source  FRB121102 was robustly identified with its host galaxy, and the distance (redshift) to it was precisely measured, interesting constraints on the photon mass were obtained \cite{2017PhLB..768..326B}. These authors derived the limit $m_\gamma<2.2 \times 10^{-14}$~eV, which corresponds to a mass of $<3.9\times 10^{-50}$~kg.

\section{Conclusions}

The FRB studies has been carried out for only 10 years. Over this time, many important properties of FRBs have been discovered, including the polarization of radio emission, repeating bursts, the identification of the host galaxy for the repeating FRB, etc. However, the FRB statistics grow quite slowly, and the origin of FRBs remains obscure. Only 'catastrophic' models in which a radio burst is accompanied by a powerful radiation in other ranges can be rejected at present. Even in this case, the hypothesis that FRB sources belong to a homogeneous population is required.

The progress in the FRB studies can be, in the first place, related to the observations at new sensitive instruments, such as UTMOST, CHIME, HIRAX, etc. Presently, the new 500-m radio telescope FAST \cite{2017RAA....17....6L} is at the commissioning stage. 
Calculations show that this instrument will be able to detect about one fast radio burst per week. About the same detection rate is expected in the future from the UTMOST telescope, and the CHIME telescope (see a recent description of this instrument in \cite{2018arXiv180311235T}), according to some estimates, can detect FRBs even at a higher rate (if their number at low frequencies is sufficiently large) \cite{2016MNRAS.458..718C}. 


A fantastic detection rate of one burst per hour is expected from the future SKA radio telescope system \cite{2016arXiv160205165K, 2015aska.confE..55M}. However, this is possible only in distant future. 

Presently, successful observations on the Parkes radio telescope continues. Dedicated FRB searches are being carried out by the 300-m Arecibo radio telescope  
\cite{2015arXiv151104132C}. 
The new system Realfast designed to identify short radio transients will be soon put into operation at the Carl Jansky VLA radio telescopes \cite{2018arXiv180203084L}. In addition, the mounting of the Apertif system \cite{2017arXiv170906104M} at the WSRT radio telescope at Westerbork  (The Netherlands) will enable this instrument to start actively searching for fast radio transients. 
It is important to note that this telescope will observe the Northern sky which has not been sufficiently surveyed so far.




It is very likely that as in the case with gamma-ray bursts, the decisive role to solve the FRB problem will be played by the identification of the events in other electromagnetic bands. This could be all-sky gamma-ray observations or optical observations; in the last case, many hopes are related to the construction of the new-generation large survey telescope LSST \cite{2016ApJ...824L..18L}. 


Methods of quick FRB identification and searches for accompanying transients are being rapidly developed. For example, the project  'Deeper Wider Faster'  that includes 
more than 30 instruments operating at different energy bands as well as cosmic ray and gravitational wave detectors  is under way
 \cite{2018arXiv180201100A}. 
 A rapid coordinated real-time exchange of information after detection of a fast radio burst (for example, by the Parkes radio telescope) gives hope to obtain important new results in the nearest future.  

Anyway, so far fast cosmic radio bursts that appear each minute on the sky remain one of the most interesting unsolved problems of the modern astrophysics.   

\section*{Acknowledgements}

The authors thank the anonymous referee for useful notes. The work of S.B. Popov (Sections 2, 3.1, 4.1, 4.2, and 4.4) and M.S. Pshirkov (Sections 3.2 and 4.3.1) is supported by the RSF grant  14-12-00146. The work of K.A. Postnov is supported by the RSF grant 16-12-10519
(Sections 1.1-1.3, 4.3.2, and general editing). S.B. Popov thanks M. Lyutikov and V.S. Beskin for numerous discussions. 

\bibliographystyle{UFN}
\bibliography{frb}
\end{document}